\documentclass[%
 aip,
 amsmath,amssymb,
 reprint,%
]{revtex4-1}

\usepackage{graphicx}
\usepackage{dcolumn}
\usepackage{bm}

\usepackage[utf8]{inputenc}
\usepackage[T1]{fontenc}
\usepackage{mathptmx}
\usepackage{etoolbox}

\makeatletter
\def\@email#1#2{%
 \endgroup
 \patchcmd{\titleblock@produce}
  {\frontmatter@RRAPformat}
  {\frontmatter@RRAPformat{\produce@RRAP{*#1\href{mailto:#2}{#2}}}\frontmatter@RRAPformat}
  {}{}
}%
\makeatother
\begin{document}


\title{Analysis of magneto-optical Kerr spectra of ferrimagnetic Mn$_4$N}
\author{J. Zemen}
\affiliation{Faculty of Electrical Engineering, Czech Technical University in Prague, Technická 2, 160 00, Prague 6, Czech Republic}

\date{\today}

\begin{abstract}
Simulations of magneto-optical Kerr effect in biaxially strained Mn$_4$N are performed using density functional theory and linear response theory. We consider three ferrimagnetic phases, two collinear and one noncollinear,  which have been corroborated separately by earlier studies. The simulated spectra are compared to magneto-optical data available in recent literature. A collinear ferrimagnetic phase with a small saturation magentization, a large perpendicular anisotropy, and Curie temperature above 700~K is found to be consistent with the measured spectra. We hypothesise that an admixture of the noncollinear phase, which could explain the lower than predicted net moment and magnetic anisotropy observed experimentally, is also present. 
\end{abstract}

\maketitle

Traditionally, ferrimagnets have been applied in magneto-optical recording,\cite{chaudhari1973amorphous,carey1995magneto,hansen1989magnetic} where a laser beam is used to increase the temperature close to the Curie point so that the magnetization can be reversed by a small applied magnetic field. Ferrimagnetic alloys containing rare earth and transition metals, such as GdFeCo and TbCo, with large perpendicular anisotropy near room temperature have show exceptional performance in this area.\cite{tsunashima2001magneto}

Recently, ferrimagnetic materials have attracted much attention for applications in high-density magnetic random access memories (MRAM) and logic devices, which can combine high perpendicular magnetic anisotropy (PMA) with saturation magnetization much lower than in typical ferromagnets and with Curie temperature well above room temperature.\cite{Suemasu2021, kim2022ferrimagnetic,zhou2021efficient,siddiqui2018current}

Ferrimagnets usually consist of two magnetic sublattices with anfiterromagnetic coupling between them. This antiferromagnetic exchange interaction allows for fast spin dynamics which has motivated the extensive research in antiferromagnetic spintronics.\cite{jungwirth2016antiferromagnetic,vzelezny2018spin,baltz2018antiferromagnetic}
The absence of net magnetic moment in antifferomagnetic materials prevents dipolar coupling between closely packed memory arrays. However, at the same time, it precludes control of the magnetic domain structure by external magnetic filed. The small but finite net magnetizaiton of ferrimagnetic materials alleviates this problem. Moreover, close to a compensation point the reduced net spin polarization makes the state of a ferrimagentic layer in, e.g., a magnetic tunnel junction (MTJ), more susceptible to spin-transfer torque (STT).
\cite{kim16gyoungchoon,Gushi2019,caretta2018fast} 

The compensation of magnetic sublattices is commonly achieved in metallic alloys containing rare earths mentioned above. However, one can avoid the reliance on rare earth elements by turning to metallic antiperovskite nitrides some of which have collinear or non-collinear ferrimagnetic structure.\cite{fruchart1977magnetic,fruchart1978magnetic,fruchart1979non}
Mn-based antiperovskiet nitrides are a broad family of materials hosting a range of phenomena including magneto-transport,\cite{gurung2019anomalous,boldrin2019anomalous,johnson2022identifying} magneto-caloric,\cite{matsunami2015giant,boldrin2018multisite,boldrin2021barocaloric} or magneto-optical properties tunable by chemical composition or lattice strain. \cite{wu2013magnetic, singh2021multifunctional,boldrin2019biaxial} We predicted that Mn$_3$GaN, which is a widely studied member of this family with a fully compensated triangular antifferomagntic ground state (cubic lattice), can develop a collinear ferrimagnetic (FIM) phase when applying compressive biaxial strain at room temperature.\cite{zemen2017frustrated} Subsequently, we used magneto-optical Kerr spectroscopy to analyse the magnetic structure of Mn$_3$NiN thin film which is a closely related antiperovskite with triangular antiferromagnetic ground state. The measured data are consistent with the presence of a collinear FIM phase at room temperature.\cite{johnson2023room}

Mn$_4$N is another member of the antiperovskte family. Its magnetic structure is even more complicated than that of Mn$_3$NiN as Ni on the $1a$ site is replaced by another Mn with a large magnetic moment. Mn$_4$N has received much attention in recent years mainly due to a large perpendicular magnetic anisotropy (PMA $\approx 10^5$ J/m$^3$) and ultrafast response to external field.\cite{Yasutomi2014,Shen2014,ito2016perpendicular,gushi2019large,Hirose2020,Zhou2021,Isogami2022, Li2022a,Zhang2023, Imamura2023}
In addition, the saturation magnetization, $M_s$ is relatively low as demonstrated by numerous studies of Mn$_4$N thin films listed in Table~\ref{table_PMA}.
Therefore, low critical STT-switching current density, $J_c \propto \alpha M_s t H_k$ (where $\alpha$ is the damping constant, $t$ is the magentic layer thickness, and $H_k$ is the anisotropy field proportional to the PMA energy density, $K_u$) is expected.\cite{Isogami2020} Moreover, the shape anisotropy is negligible due to low $M_s$ so the magnetic anisotropy is dominated by the magnetocrystalline contribution.\cite{Isogami2020} 

In order to utilize the potential of Mn$_4$N for spintronic applications such as MTJs or skyrmionic devices,\cite{Bayaraa2021} it is crucial to attain thorough understanding of the microscopic origin of the large PMA combined with low magnetization and to be able to grow thin films with well defined magnetic structure on substrates compatible with CMOS technology. (So far, superior magnetic properties including sharp magnetization switching have been observed on STO substrates,\cite{Gushi2018} which complicates integration due to the requirement of post-growth annealing at high temperatures to achieve crystallinity of the STO barrier.)

\begin{table}
\caption{Comparison of Mn$_4$N films with thickness ($t$), and measured magnetic anisotropy energy (K$_u$) and saturation magnetization (M$_s$).}
\label{table_PMA}
\begin{tabular}{|c|c|c|c|c|c|} 
\hline
Substrate & $c/a$ & $t$ [nm]  &  method    &   K$_{u}$ [MJ/m$^3$] & M$_s$ [kA/m] \\
 \hline \hline
MgO\cite{Shen2014} & 0.987 & 35 & PLD & 0.16 & 157 \\
\hline
MgO\cite{Yasutomi2014} & 0.99 & 26 & MBE & 0.22 & 145 \\
\hline
MgO\cite{Kabara2015} & 0.99 & 100 & Sputtering & 0.88 & 110 \\
\hline
STO\cite{ito2016perpendicular} & 0.99 & 25 & MBE & 0.1 & 80 \\
\hline
MgO\cite{Foley2017} & 0.983 & 9 & MBE & 0.18 & 127 \\
\hline
MgO\cite{Anzai2018} & 0.991 & 30 & MBE  & 0.075 & 100 \\
\hline
MgO\cite{Gushi2018} & - & 10 & MBE & 0.11 & 118 \\
\hline
STO\cite{Gushi2018} & - & 10 & MBE & 0.11 & 105 \\
\hline
STO\cite{gushi2019large} & - & 10 & MBE & 0.11 & 71 \\
\hline
MgO\cite{Hirose2020} & 0.993 & 18 & MBE & 0.06 & 63 \\
\hline
STO\cite{Hirose2020} & 0.989 & 17 & MBE & 0.126 & 73 \\
\hline
LAO\cite{Hirose2020} & 0.998 & 19 & MBE & 0.045 & 53 \\
\hline
MgO\cite{Isogami2020} & 0.99 & 30 & Sputtering & 0.1 & 80 \\
\hline
MgO\cite{He2022} & 0.99 & 28 & Sputtering & 0.17 & 156 \\
\hline
Glass\cite{Li2022a} & 0.993 & 45 & Sputtering & 0.022 & 36 \\
\hline
Glass\cite{Li2022} & 0.988 & 48 & Sputtering & 0.073 & 99 \\
\hline
MgO/VN\cite{Imamura2023} & 0.987 & 28 & Sputtering & 0.043 & 85 \\
\hline
\end{tabular}
\end{table}

In studies listed in Table~\ref{table_PMA}, Mn$_4$N has been deposited since 2014 on a range of substrates with different lattice mismatches, e.g., MgO, SrTiO$_3$ (STO), and LaAlO$_3$ (LAO) with mismatch of approximately $-6\%$, $-0.1\%$, and $+2\%$, respectively, assuming (001) surfaces and lattice constant of cubic Mn$_4$N equal to 0.3865 nm.\cite{takei1962magnetic} 
It has been observed using x-ray diffraction that the films have $c/a \approx 0.98-0.99$, where $a$ and $c$ are the in-plane and out-of-plane lattice constant, respectively, despite the different magnitude and sign of lattice mismatch. 

\begin{figure}
\includegraphics[width=0.97\columnwidth]{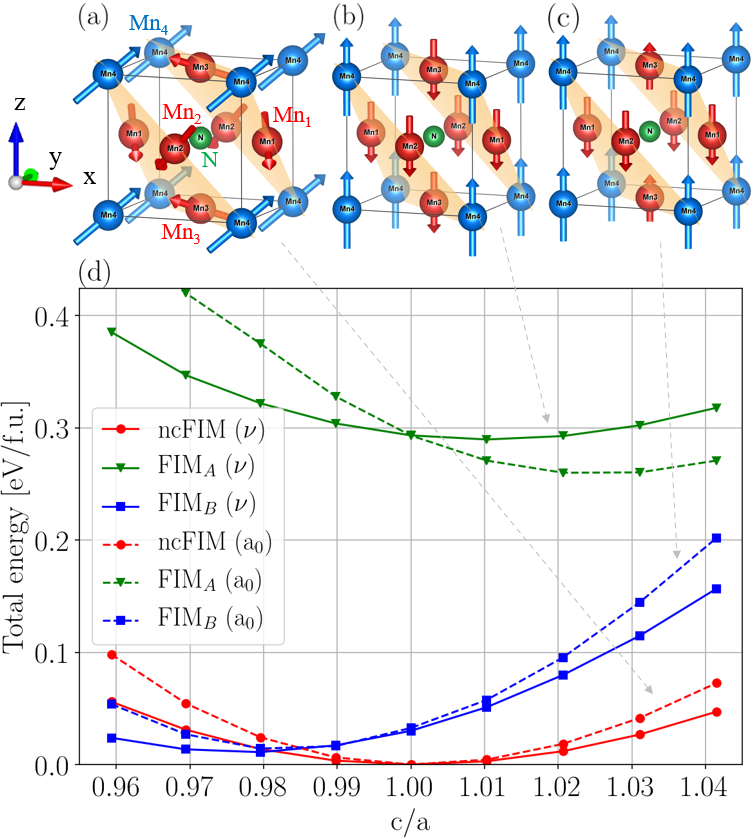}
\caption{Comparison of ferrimagnetic structures considered: (a-c) unit cells with magnetic moments and chemical identities: Mn on $3c$ sites in red, Mn on $1a$ sites in blue, N in green, and (d) corresponding calculated total energies vs $c/a$ ratio.}
\label{fig:toten}
\end{figure}

Therefore, experimental studies that keep track of $c/a$ and PMA generally conclude that the origin of PMA in Mn$_4$N films is the tetragonal distortion.\cite{Hirose2020,Yasutomi2014,Shen2014,Kabara2015} 
Moreover, studies that include $ab~initio$ simulations have associated the observed PMA with a collinear ferrimagnetic phase, so called FIM$_B$ phase, with total energy minimum at $c/a=0.98$.\cite{ito2016perpendicular,Isogami2020}
Collinear ferrimagnetic phases FIM$_A$ and FIM$_B$ shown in Fig.~\ref{fig:toten} were revealed by neutron diffraction experiments as early as 1962.\cite{takei1962magnetic}
Ito et al.\cite{ito2016perpendicular} and Isogami et al.\cite{Isogami2020} demonstrated by $ab~inito$ simulations that FIM$_B$ has a significantly lower total energy than FIM$_A$ in the range $c/a \in (0.96-1.1)$, in agreement with our simulations shown in Fig.~\ref{fig:toten}(d).
Both theoretical studies suggest that this intrinsic tetragonal phase with large PMA explains why $c/a \in (0.98-0.99)$ has been reported in Mn$_4$N films epitaxially grown on different substrates regardless of the film thickness and lattice mismatch.

However, negligible dependence of PMA on $c/a$ was predicted by Isogami et al.\cite{Isogami2020} which is in disagreement with PMA measured by Hirose et al.\cite{Hirose2020} in Mn$_4$N on MgO, STO, and LAO. 
Moreover, the theoretical studies mentioned so far\cite{ito2016perpendicular,Isogami2020} have not considered any noncollinear magnetic phases despite the fact that a noncollinear ferrimagnetic phase (ncFIM) was identified by neutron diffraction in bulk Mn$_4$N in 1979.\cite{fruchart1979non} 
Magnetic moments of Mn atoms in this "umbrella-like" structure shown in Fig.\ref{fig:toten}(a) do not mutually compensate as in cubic Mn$_3$NiN or Mn$_3$GaN even though the lattice has cubic symmetry with space group $Pm\bar{3}m$.\cite{fruchart1979non} This is due to the fact that $1a$ corner site is occupied by Mn with a large local magnetic moment, $m_{1a}$. The moments in face-center $3c$ positions, $m_{3c}$, are tilted out of the (111) planes (where the Mn atoms form a kagome lattice) by approximately 20$^{\circ}$ to have a component along the [111] axis, antiparallel to $m_{1a}$. The ncFIM structure was confirmed computationally by Uhl et al.\cite{Uhl1997} and more recently by Zhang et al.\cite{Zhang2022a} including the net moment along the [111] axis, $m_{net} = 1.1\mu_B$. 
The same team also proposed related noncollinear ferrimagnetic phases in Mn$_4$N film on MgO\cite{He2022} based on ncFIM and the coplanar triangular antiferromatic structure of Mn-based antiperovskite nitrides.

Here we compare the collinear and noncollinear FIM phases of strained Mn$_4$N based on total energy and Magneto-optical Kerr effect (MOKE) calculated using Density Functional Theory (DFT) and linear response theory. This is primarily motivated by a recent study of MOKE in 23~nm thick Mn$_4$N films sputtered on MgO substrate.\cite{Sakaguchi2023} The authors interpret their MOKE spectra based on projected densities of states (PDOS) simulated by Isogami et al.\cite{Isogami2020} and conclude that: "The fine structures observed in Kerr rotation could be attributed to a superposition of different magnetic phases from the dominant ferrimagnetism in Mn$_4$N, although theoretical calculations may be necessary for further interpretation." Moreover, an in-plane component of magnetization has been detected in one Mn$_4$N film deposited on MgO by MBE,\cite{wang2017magnetic} and the authors explained it by presence of a FIM phase with magnetization along the [111] axis.\cite{Takei1960,fruchart1979non}

Therefore, we simulate the MOKE spectra in this work for the three main magnetic structures FIM$_A$, FIM$_B$, and ncFIM.
We employ noncollinear spin polarized DFT following the approach of Ref.~[\onlinecite{feng2015large}] and our earlier work.\cite{johnson2023room} We use the projector augmented wave method as implemented in the VASP code\cite{kresse1993ab} with with generalized gradient approximation (GGA) parameterized by Perdew–Burke–Ernzerhof.\cite{perdew1996generalized} Our results were obtained using a 500~eV energy cutoff and a 23 × 23 × 23 k-mesh (for a unit cell with 5 atoms) to ensure convergence (in agreement with numerical settings of Ref.~[\onlinecite{Isogami2020}]). The valence configurations of manganese and nitrogen are 3$d^6$4$s^1$ and 2$s^2$2$p^3$, respectively.

\begin{table}
\caption{Comparison of magnetic space groups and corresponding anomalous Hall conductivity tensor (AHC) for each pahse.}
\label{table_AHC}
\begin{tabular}{|c|c|c|} 
\hline
   &            ncFIM          &           FIM$_A$, FIM$_B$           \\ 
   \hline \hline
Space group   &      123, P4/mmm                &     123, P4/mmm        \\ 
\hline
Mag. space group   &      12.62, C2'/m'  &     123.345, P4/mm'm'         \\ 
\hline
 AHC, $\sigma_{\alpha, \beta}$       &             $ \left( \begin{array}{ccc} 0  & \sigma_{xy} & \sigma_{xz}   \\ -\sigma_{xy} & 0 & \sigma_{xz} \\ -\sigma_{xz} & -\sigma_{xz} & 0 \end{array}\right)$          &  $\left( \begin{array}{ccc} 0  & \sigma_{xy} & 0   \\ -\sigma_{xy} & 0 & 0 \\ 0 & 0 & 0 \end{array}\right)$                     \\ 
\hline 



\end{tabular}
\end{table}

Taking into account our recent study of MOKE spectra in the closely related antiperovskite Mn$_3$NiN,\cite{johnson2023room} and earlier MOKE studies of some collinear antiferromagnets such as CuMnAs,\cite{veis2018band} we modify the intra-atomic Coulomb interaction within GGA through the rotationally invariant approach to GGA+U as proposed by Dudarev et al.\cite{dudarev1998electron} We explore values of U from 0.7~eV (refined in Ref.~[\cite{johnson2023room}]) to 2.2~eV on the Mn-3$d$ orbital. (All data shown here in figures were simulated using U = 0.7~eV.) This repulsion lifts the unoccupied manganese $d$-states further away from the Fermi level, resulting in a blueshift in the optical and magneto-optical responses which improves the agreement with the available measured data.\cite{Sakaguchi2023} 

The unit cells and corresponding total energies as functions of the $c/a$ ratio are shown in Fig.~\ref{fig:toten}. By analysing available x-ray data of films listed in Table~\ref{table_PMA} we noticed a range of values of Poisson's ratio, $\nu$. Therefore, for each $c/a$ ratio, we calculate the total energy assuming $\nu=0.33$\cite{He2022} as well as $a=a_0$, where $a_0=0.389$~nm is the equilibrium lattice parameter for ncFIM from our DFT simulations. This is close to the experimental value $a_0=0.3865$~nm and to an earlier DFT calculation, $a_0=0.382$~nm.\cite{Zhang2022a} We note that our conclusion is independent of this choice:  
The energy minimum for FIM$_A$ phase obtained at $c/a>1$ is more than 0.2~eV higher than the energy minimum of FIM$_B$ at $c/a=0.98$, in agreement with earlier DFT studies,\cite{ito2016perpendicular,Isogami2020} which suggest that the Mn-Mn direct AFM interaction might be stabilizing the FIM$_B$ structure.
However, the ground state of ncFIM phase ($c/a=1$) is another 30~meV lower than the energy minimum of FIM$_B$. (This energy difference is much bigger than 4~meV between ncFIM and a cubic collinear FIM phase with $m_{1a}$ ($m_{3c}$) parallel (antiparallel) to the [111] axis predicted by Zhang et al.\cite{Zhang2022a})

Therefore, it is conceivable that ncFIM phase with $c/a \in (0.99-1.0)$ coexists with FIM$_B$ phase in films where the lattice mismatch with the substrate does not induce a large biaxial strain thanks to dislocations\cite{Shen2014,ito2016perpendicular} or a "dead layer"\cite{Isogami2020} immediately above the interface. Areas of the film with less dislocations (better epitaxy) are then more likely to stabilise the tetragonal FIM$_B$ phase. This suggestion further elaborates on the aforementioned explanation why Mn$_4$N films with $c/a \approx 0.99$ have been reported on various substrates regardless of the film thickness and lattice mismatch.
Furthermore, a mixture of FIM$_B$ and ncFIM phases could explain the lower-than-predicted M$_s$ and PMA as we will discuss below.

\begin{figure}
\includegraphics[width=0.97\columnwidth]{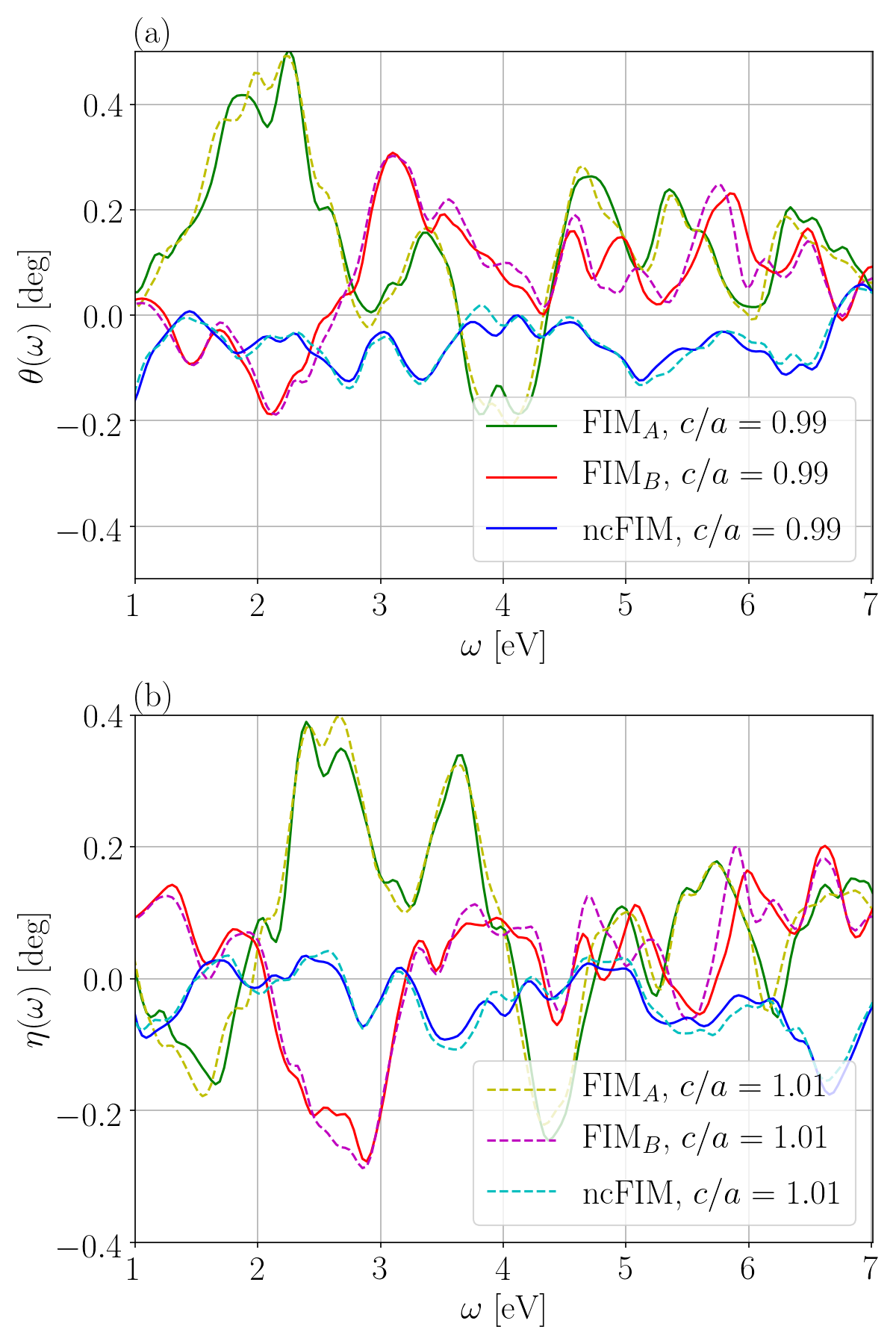}
\caption{Simulated MOKE spectra: (a) Kerr rotation and (b) Kerr ellipticity; for two values of $c/a$ and U =~0.7~eV (without smoothing).}
\label{fig:moke}
\end{figure}

MOKE spectra offer a valuable insight into the magnetic structure of thin films so we take this opportunity to compare spectra simulated for all there ferrimagnetic phases to the measured data.\cite{Sakaguchi2023}
As in our recent study of MOKE spectra in Mn$_3$NiN,\cite{johnson2023room} we note that MOKE is an optical counterpart of the Anomalous Hall Effect (AHE). Both MOKE and the intrinsic contribution to Anomalous Hall Conductivity (AHC), $\sigma_{\alpha,\beta}$, originate (within linear response thoery) in non-vanishing integral of Berry curvature $\Omega_{n,\alpha,\beta}(k)$ over the Brillouin zone:

\begin{equation}
\label{eq:ahc}
\begin{aligned}
\sigma_{\alpha, \beta} & =\frac{-e^2}{\hbar} \int \frac{d k}{2 \pi^3} \sum_{n \text { (occ. })} f[\varepsilon_n(k)-\mu] \Omega_{n, \alpha, \beta}(k), \\
\Omega_{n, \alpha, \beta}(k) & =-2 I \sum_{m \neq n} \frac{\left\langle k m\left|\nu\alpha(k)\right| k n\right\rangle\left\langle k n\left|\nu\beta(k)\right| k m\right\rangle}{\left[\varepsilon_{k n}-\varepsilon_{k m}\right]^2},
\end{aligned}
\end{equation}
where  $f[\varepsilon_n(k)-\mu]$ denotes the Fermi distribution function
with $\varepsilon_n(k)$ and $\mu$ being the energy eigenvalues of occupied (unoccupied) Bloch band, $n$, and the Fermi energy, respectively, and where $\nu\alpha(k)$ corresponds to the velocity operator in Cartesian coordinates.

The Kerr angle ($\theta_K$) and ellipticity ($\eta_K$) in case of polar-MOKE geometry depend on AHC as follows\cite{johnson2023room}:

\begin{equation}
\label{eq:kerr}
\theta_K+i \eta_K=\frac{-\sigma_{x y}}{\sigma_{x x} \sqrt{1+i(4 \pi / \omega) \sigma_{x x}}},
\end{equation}
where $\omega$ is the photon energy and we assume that the z-axis (parallel to the incident beam) is perpendicular to the film surface.

The presence of MOKE and AHE can be determined by analyzing the transformation properties of the Berry curvature under all symmetry operations of a particular magnetic space group. In Table~\ref{table_AHC} we list the space groups of the three FIM phases of Mn$_4$N (subject to tetragonal distortion) obtained by FINDSYM software.\cite{stokes2005findsym,findsym}
The last row of Table~\ref{table_AHC} presents the form of the AHC tensor in linear response regime obtained using software Symmetr\cite{symmetr} considering both set of symmetry operations and the spin-orbit coupling.
We note that both collinear FIM phases share the same form of AHC tensor with one independent nonzero element, $\sigma_{x,y}$, inducing the polar-MOKE.
The AHC tensor of ncFIM has two independent nonzero elements, $\sigma_{x,y}$ and $\sigma_{x,z} = \sigma_{y,z}$ as in case of strained Mn$_3$NiN\cite{johnson2023room}. Cubic Mn$_4$N would have a magnetic space group 166.101, R$\bar{3}$m' and $\sigma_{x,y} = \sigma_{x,z} = \sigma_{y,z}$. The listed forms of the AHC tensor are determined by the symmetry of the structure rather than the net magnetization so they would not change (AHE and MOKE would not vanish) even if the net magnetic moment vanished, i.e., if full compensation of the ferrimagnet was achieved, which is desirable when seeking ultrafast spintronic devices.\cite{Bayaraa2021,Zhang2022a}

Fig~\ref{fig:moke} presents the main result of the work. It shows Kerr rotation and Kerr ellipticity as a function of energy, $\omega \in (1-7)$~eV. Our model does not include the intraband contribution which dominates below 1~eV, hence the choice of energy interval. We calculated the spectra for several ratios $c/a \in (0.97-1.03)$ but the dependence on tetragonal distortion appears to be much smaller than the differences between the three FIM phases so we plot only spectra for $c/a=0.99$ and 1.01. The spectra in this figure include fine features as we use small Gaussian smearing, $\sigma=0.01$~eV, to treat the partial occupancies in k-space integration. This approach would correspond to experimental data measured in a crystalline film with few defects at low temperature. However, the available MOKE data were measured in sputtered films at ambient temperature.\cite{Sakaguchi2023} Therefore, we include also Fig.~\ref{fig:mokesmeared} where the fine features are smoothed out.

\begin{figure}
\includegraphics[width=0.97\columnwidth]{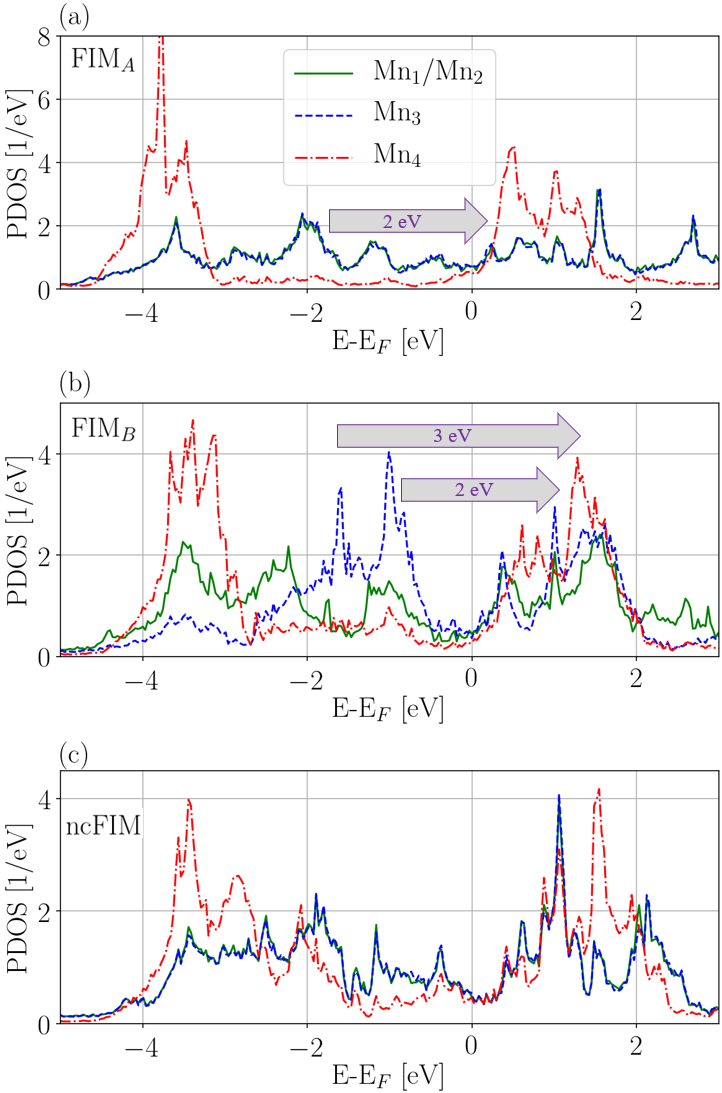}
\caption{Comparison of Projected Density of States (PDOS) for FIM$_A$ (a), FIM$_B$ (b), and ncFIM (c) phases with $c/a=0.99$ and U =~0.7~eV. Grey arrows indicate potentially dominant excitations.}
\label{fig:pdos}
\end{figure}

In order to interpret our spectra based on features of the band structure, we plot the projected DOS for all three phases in Fig.~\ref{fig:pdos}. We resolve PDOS only for Mn 3d orbitals as the other contributions are small and too far from the Fermi energy to play an important role in the visible magneto-optical response. Mn$_1$, Mn$_2$, and Mn$_3$ occupy the $3c$ sites with cartesian coordinates (0.0,~0.5,~0.5), (0.5,~0.0,~0.5), and (0.5,~0.5,~0.0), respectively, whereas Mn$_4$ occupies the $1a$ site with coordinates (0,~0,~0) as shown in Fig.~\ref{fig:toten}. We note that our PDOS for FIM$_B$ phase is in agreement with Fig.~7(a) of Ref.~[\onlinecite{Isogami2020}].

Figure~\ref{fig:pdos}(a) for FIM$_A$ shows PDOS with one dominant transition indicated by a grey arrow between a peak in occupied states of Mn$_{1-3}$ and a peak in excited states of Mn$_4$. This transition described by energy difference, $dE \approx 2$~eV, corresponds to the sharp peak in magneto-optical response at $\omega \approx 2$~eV. 
Figure~\ref{fig:pdos}(b) for FIM$_B$ shows PDOS with two dominant transition indicated again by arrows described by $dE \approx 2$~eV and 3~eV which correspond to a dip and a peak in Kerr rotation, respectively.

Figure~\ref{fig:pdos}(c) for ncFIM shows PDOS lacking prominent peaks around 2~eV below the Fermi level, which are present in case of FIM$_B$. The peak in PDOS of Mn$_4$ present in FIM$_A$ around 0.5~eV above Fermi energy is shifted to 1~eV in ncFIM. Such band structure results in absence of prominent peaks in the magneto-optical response at photon energies below 4~eV. We conclude that the predicted spectral features can be interpreted based on transitions from Mn-3d orbitals on $3c$ sites to Mn-3$d$ orbitals on $1a$ site and that the three FIM phases should relatively easily distinguishable even after smearing their distinctive features by effects such as lattice defects or elevated temperature.

\begin{figure}
\includegraphics[width=0.97\columnwidth]{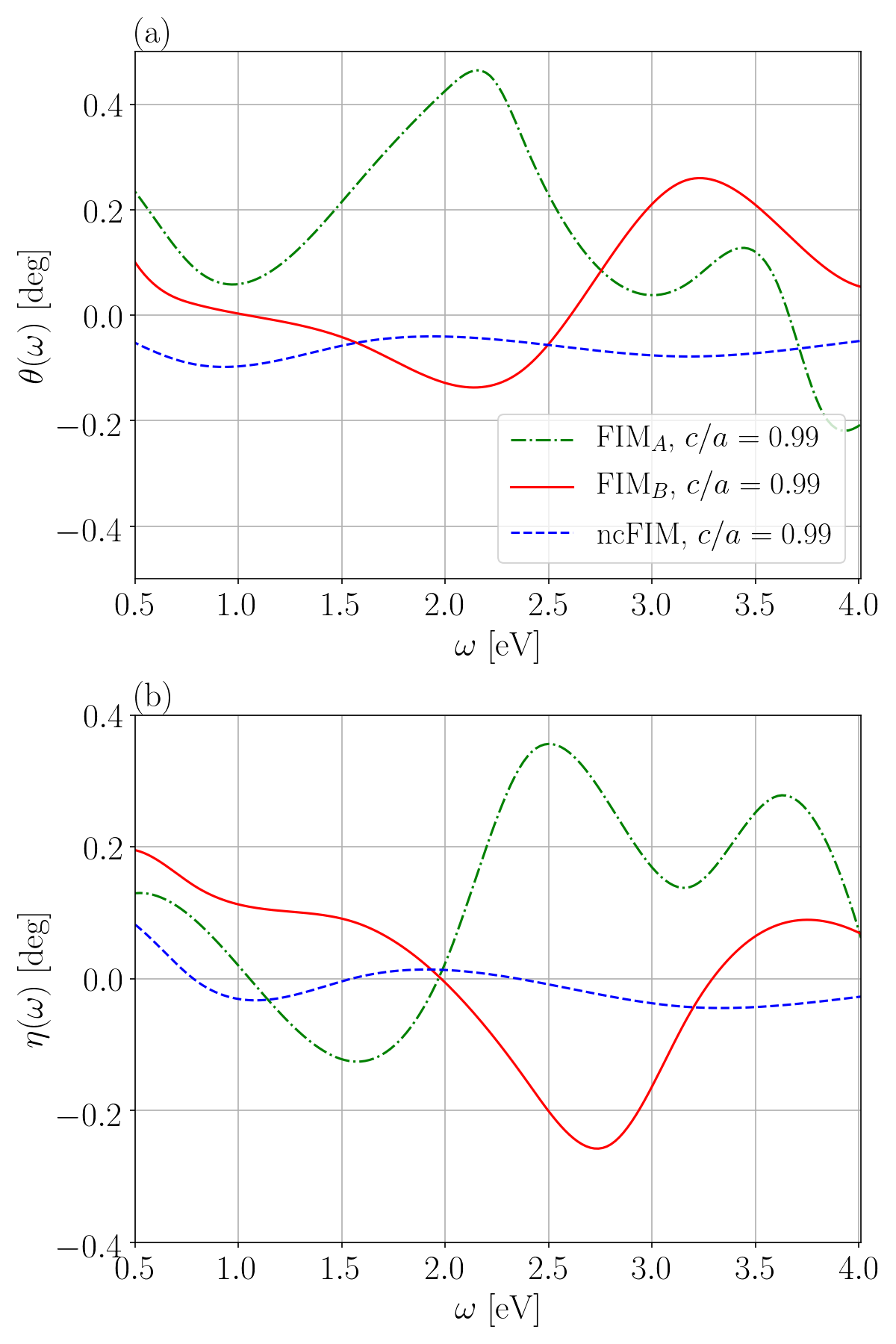}
\caption{Simulated MOKE spectra: (a) Kerr rotation and (b) Kerr ellipticity; for $c/a=0.99$, U =~0.7~eV, and smoothing of DFT data in Fig.~\ref{fig:moke}.}
\label{fig:mokesmeared}
\end{figure}

In order to compare our data to MOKE spectra measured by Sakaguchi et al.,\cite{Sakaguchi2023} we interpolate our curves plotted in Fig.~\ref{fig:moke} using the UnivariateSpline function from Python Scipy package with smoothing parameter set to 0.0008 on a linear grid with 200 points. The smeared spectra for $c/a=0.99$ are shown in Fig.~\ref{fig:mokesmeared}. Our Kerr rotation, $\phi(\omega)$, and ellipticity, $\eta(\omega)$, can be compared directly to Fig.~5(a) and (b) of Ref.~[\onlinecite{Sakaguchi2023}], respectively, where the boron content in Mn$_4$N is zero. 

Firstly, we note the the amplitude of the measured spectra is approximately 0.1~degree which is an order of magnitude larger than in the related noncollinear Mn-based systems such as Mn$_3$NiN\cite{johnson2023room} and Mn$_Sn$.\cite{higo2018large} The amplitude of the simulated spectra is only a factor of two larger which is an unexpectedly good agreement indicating high quality of the Mn$_4$N film.

Secondly, the measured $\phi(\omega)$ is positive below 1.5~eV, has a dip around 2.2~eV, and seems to come back to positive values above 3~eV, where the studied interval ends. Such trend is observed only in data simulated for FIM$_B$ phase, although the crossing points, $\phi(\omega)=0$, are shifted to lower energies (1 and 2.6~eV instead of the measured 1.5 and 3~eV). A shift of optical and magneto-optical responses is commonly achieved in DFT studied by increasing the Coulomb repulsion on sites with more localized states (here the Mn-3$d$ states). Therefore, we simulated the spectra for a range of Hubbard parameters, $U$, from $U=0.7$~eV (shown here) to $U=2.2$~eV. Spectra with $U\approx2.2$~eV have crossing points very close to 1.5 and 3~eV, as in the experiment. However, we believe that this value of $U$ is too high compared to Mn$_3$NiN\cite{johnson2023room} and CuMnAs\cite{veis2018band} where $U=0.7$ and $U=1.7$~eV was used, respectively, for Mn-3$d$ orbitals to predict MOKE spectra in semiquantitative agreement with measurement. 

Thirdly, we speculate that the crossing point, $\phi(\omega)=0$, at 2.6~eV for FIM$_B$ could be shifted to higher energy by a assuming an admixture of ncFIM, which is negative throughout the energy interval. Such superposition of spectra would also shift the crossing point at 1~eV to lower energies in contrast to experiment. However, our predictions are less reliable below 1.5~eV as our model does not include the intraband contributions (the Drude peak).

Finally, we check the agreement in case of Kerr ellipticity, $\eta(\omega)$, which is Kramers-Kronig-related to $\phi(\omega)$.
Fig.~5(b) of Ref.~[\onlinecite{Sakaguchi2023}] shows a monotonous decrease of $\eta(\omega)$ to zero around 3~eV. As expected, FIM$_B$ is the only phase that shows such trend in the simulated spectrum but the crossing point is again shifted to lower energy.

\begin{figure}
\includegraphics[width=0.97\columnwidth]{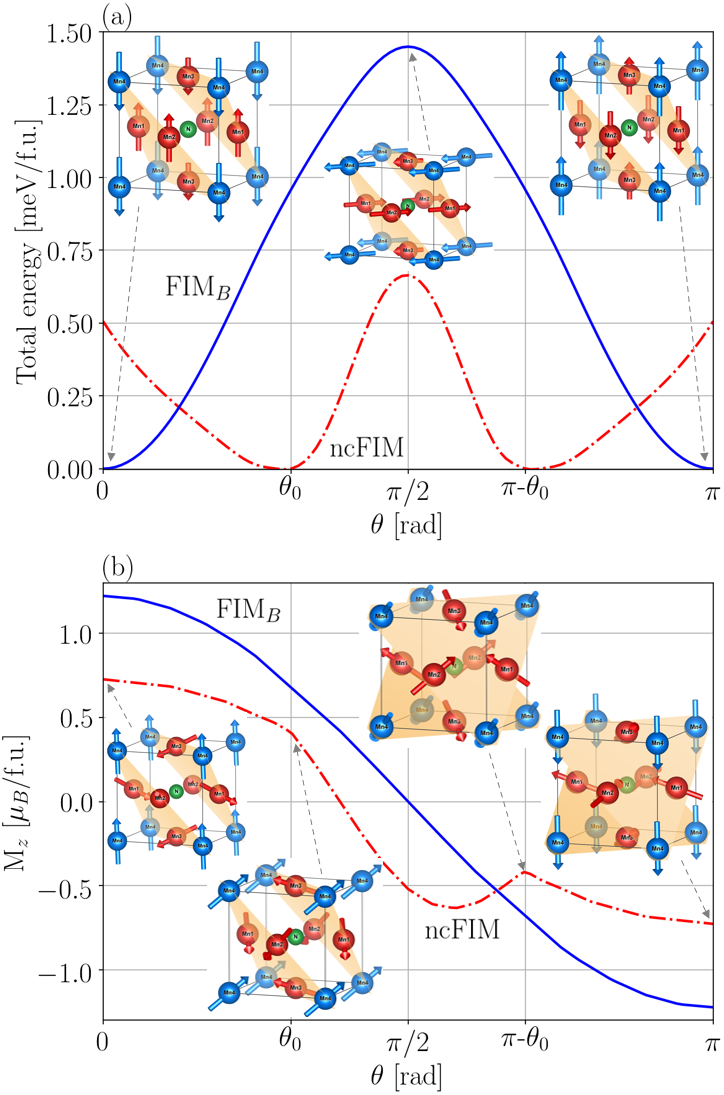}
\caption{Comparison of FIM$_B$ and ncFIM phases based on: (a) total energy and (b) magnetization projection, M$_z$; angle $\theta$ describes the switching from magnetization along the [001] axis to the opposite direction. The insets show the magnetic structure at significant points along the selected switching paths.}
\label{fig:mae}
\end{figure}

To complete the analysis and to formulate our hypothesis about dominant FIM$_B$ phase with an admixture of ncFIM phase, we discuss Figure~\ref{fig:mae} which shows the magneto-crystalline anisotropy profile and the component of magnetization perpendicular to the film  (direction of the applied field), $M_z$, for the two relevant phases. In Fig.~\ref{fig:mae}(a) the total energy is plotted as a function of angle $\theta$ between the net magnetization and the [001] axis (perpendicular to film). The insets show the orientation of the local moments of FIM$_B$ phase for net magnetization pointing along [001], [110], and [00$\overline{1}$]. This choice is relevant for the polar-MOKE experiment, where the sample is measured in magnetic field applied parallel and antiparallel to the [001] axis and the two spectra are subtracted to eliminate nonlinear MOKE effects.\cite{Sakaguchi2023} The film has to undergo a rotation of magnetic moments driven by the reversal of the perpendicular applied field between the two measurements. So it has to overcome the energy barrier of the in-plane orientation, $\theta= \pi/2$, which is $dE = 1.4$~meV per formula unit or 3.78~MJ/m$^3$ for the FIM$_B$ at $c/a=0.99$. This PMA is in perfect agreement with earlier calculations.\cite{Isogami2020} However, the value is much larger than experimental PMA $\approx 0.1$~MJ/m$^3$ as listed in Table~\ref{table_PMA}. We cannot compare PMA directly in case of Sakaguchi et al.\cite{Sakaguchi2023} as they give only the anisotropy field, $H_k=1.5$~T, which is a typical value on Mn$_4$ on MgO but we do not know the saturation magnetization and the size of the applied field. So we have to assume that the sample was fully aligned with the applied field during MOKE measurement. However, the discrepancy between the predicted and measured PMA remains an open question which complicates the interpretation of MOKE spectra.

The energy barrier is lower for ncFIM at $c/a=0.99$, $dE=0.67$~eV/f.u. or 1.81~MJ/m$^3$, which speaks in favour of our hypothesis of ncFIM and FIM$_B$ coexistence. However, the spin reorientation mechanism becomes much more complicated in case of ncFIM. There are 8 variants of this phase with the net magnetization pointing parallel or antiparallel to the 4 body-diagonals, in perfect analogy to Mn$_3$NiN.\cite{johnson2022identifying} The applied field can align the net magnetization with the [001] axis but the rotation of the moments to the opposite field orientation can go through different energy minima and energy barriers depending on the local conditions in the film. We have carried out an extensive DFT study of the total energy landscape as a function of coherent rotations of the 4 local moments. 

We considered all rotations that belong to the $Pm\bar{3}m$ space group of the cubic perovskite lattice: 2-fold and 4-fold rotations about the main axes, 3-fold rotations about the body diagonals, and 2-fold rotations about the side diagonals. Details about our findings will be summarised elsewhere. Here we show an example of a rotation between two energy minima (from [111] to [$\bar{1}$1$\bar{1}$]) which incurs the lowest energy barrier, $dE=0.67$~eV/f.u. mentioned above. This rotation is a simultaneous rotation by $\pi/2$ about the [010] axis, by $\pi/2$ about the [001] axis and by [$2\pi/3$] about [$\bar{1}$1$\bar{1}$] axis. (An intuitive simple rotation of magnetization from [111] to [11$\bar{1}$] about the [$\bar{1}$10] axis would not restore the ground state magnetic structure.)
The dash-dotted line in Fig.~\ref{fig:mae} consists of: (i) a simple rotation about the [$\bar{1}$10] axis from a state with net magnetization along [001] to the ground state along [111] denoted by $\theta_0$, (ii) the simultaneous rotation to [$\bar{1}$1$\bar{1}$] denoted by $\pi-\theta_0$), (iii) a simple rotation about [110] to a state with net magnetization along [00$\bar{1}$]. All four significant states are depicted as insets in Fig.~\ref{fig:mae}(b).
Notably, the structure in the first inset is of the same type (and has the same direction of magnetization) as the noncollinear structure proposed for Mn$_4$N in Fig.~2(d) of Ref.~[\onlinecite{He2022}].

As illustrated by Fig.~\ref{fig:mae}, a film in ncFIM phase with cubic lattice or with slight tetragonal distortion, $c/a \in (0.99,1)$, can be in a multi-domain state even at saturation (when the net magnetization is fully aligned with applied field) and the magnetization reversal can follow different paths for each domain. Investigation of magnetic domain wall (WD) propagation in case of ncFIM is beyond the scope of this study but we believe that the availability of more energy minima and lower energy barriers between them (compared to FIM$_B$) would facilitate DW propagation, thereby lowering the effective anisotropy field to values observed experimentally. 

Our hypothesis would have implications for the observed saturation magnetization, $M_s$, so we include Fig.~\ref{fig:mae}(b) to show the out-of-plane magnetization component, M$_z$. Isogami et al.\cite{Isogami2020} predict 180~mT for FIM$_B$ and observe 110~mT experimentally. They are able to attribute the discrepancy to a dead layer at the interface with substrate, and nitrogen deficiency and top surface oxidation. Here we suggest that the lower $M_s$ could be due to the admixture of ncFIM with M$_s = 0.727 \mu_b$/f.u. = 143~mT. However, we admit that Zhang et al.\cite{Zhang2022a} predict $M_s = 1.24 \mu_B$/f.u. = 244~mT for ncFIM using DFT. Our M$_s$ is lower due to the use of U = 0.7~eV which leads to larger local moments on $3c$ sites and consequently more compensation of $m_{1a}$.





In summary, we simulated the MOKE spectrum of strained Mn$_4$N using DFT+U assuming three ferrimagnetic phases discovered by earlier neutron diffraction as well as theoretical studies. We compared our results to polar-MOKE spectra measured by Sakaguchi et al.\cite{Sakaguchi2023} in Mn$_4$N films on MgO substrate. We found that the key features of the simulated spectra are consistent with the measured spectrum only in case of the FIM$_B$ phase. The agreement of the simulated Kerr rotation could be further improved if a fraction of the ncFIM phase was added to the dominant FIM$_B$ phase. At the same time, the admixture of ncFIM could explain the lower PMA and M$_s$ observed experimentally. We believe that our analysis will motivate further MOKE studies where the applied field can be inverted along a chosen path, using a vector magnet in particular. This could shed more light on the ncFIM phase preferred at lower tensile strains and enable sub-nanosecond spin dynamics at room temperature. 

\begin{acknowledgments}
We acknowledge fruitful discussions with Freya Johnson, Lesley F. Cohen, Martin Veis, and Jakub Železný.
This work was supported by the Ministry of Education, Youth and Sports of the Czech Republic through the e-INFRA CZ (ID:90254)
\end{acknowledgments}

\section*{Data Availability Statement}

The data that support the findings of this study are available from the
corresponding author upon reasonable request.

\appendix

\bibliography{Mn4N_APL}

\end{document}